\documentclass{article}
\usepackage{spconf,amsmath,graphicx}
\usepackage{threeparttable}
\usepackage{multirow}
\usepackage{amssymb}
\usepackage{epsfig}
\usepackage{epstopdf}
\usepackage{etoolbox}
\usepackage{adjustbox}
\usepackage{caption}
\usepackage{subcaption}
\usepackage{graphicx}
\usepackage{enumitem}

\title{EEG-BASED VIDEO IDENTIFICATION USING GRAPH SIGNAL MODELING AND GRAPH CONVOLUTIONAL NEURAL NETWORK}
\name{Soobeom Jang, Seong-Eun Moon and Jong-Seok Lee \thanks{\textcopyright 2018 IEEE. Personal use of this material is permitted. Permission from IEEE must be obtained for all other uses, in any current or future media, including reprinting/republishing this material for advertising or promotional purposes, creating new collective works, for resale or redistribution to servers or lists, or reuse of any copyrighted component of this work in other works.} \thanks{
This work was supported by Basic Science Research Program through the National Research Foundation of Korea (NRF) funded by the Korea government (MSIT) (NRF-2016R1E1A1A01943283).} 
}
\address{School of Integrated Technology, Yonsei University, South Korea \\
    email:\{soobeom.jang, se.moon, jong-seok.lee\}@yonsei.ac.kr
}
\begin{document}
%\ninept
%
\maketitle
\begin{abstract}
This paper proposes a novel graph signal-based deep learning method for electroencephalography (EEG) and its application to EEG-based video identification. We present new methods to effectively represent EEG data as signals on graphs, and learn them using graph convolutional neural networks. Experimental results for video identification using EEG responses obtained while watching videos show the effectiveness of the proposed approach in comparison to existing methods. Effective schemes for graph signal representation of EEG are also discussed.
\end{abstract}
\section{Introduction}
\label{sec:introduction}

The brain signal provides the most comprehensive information regarding the mental state of a human subject. Many applications exploiting brain signals have been attempted, including neurological disease detection, emotion recognition, and behavioral modeling. There are several types of brain signals that can be used, such as electroencephalography (EEG), magnetoencephalography (MEG), and functional magnetic resonance imaging (fMRI). In particular, EEG has been considered as a promising solution for various real-world applications thanks to the advances of portable EEG devices and signal processing techniques.

Brain signals generated from different brain regions may have relationship, which can be exploited for analysis. One way to describe such relationship is based on the physical distances between different regions. Another way developed recently is based on functional connectivity, which is defined as similarity between signals from different regions, e.g., cross-correlation, mutual information, phase synchronization, and imaginary coherence \cite{van2010exploring}\cite{hassan2014eeg}.

% \cite{chen2015identifying}
Recently, graph signal processing has been proposed to process irregularly structured signals effectively \cite{Shuman2013emerging}\cite{sandryhaila2013discrete}. It is to extend traditional digital signal processing techniques to signals that are not sampled on regular domains (such as time and grid space) but reside on graphs composed of vertices and edges. Furthermore, deep learning on graph signals has been also studied, and neural network structures for graph signals were proposed \cite{Defferrard2016convolutional}\cite{Kipf2016semi}\cite{Pham2017column}.

Brain signals are good examples of graph signals, because graphs are suitable to represent physical or functional connectivity across different brain regions. However, there exist little work on applying graph signal processing techniques and graph signal-based deep learning methods to model brain signals, particularly EEG. This is probably due to the limited number of channels (i.e., electrodes) of EEG, which may not be sufficient for rich graph representation. 

This paper proposes a method for deep learning on graph signals for EEG analysis and its application to EEG-based video identification. To our best knowledge, this is the first attempt to apply graph signal-based deep learning techniques to EEG. In particular, we present various ways to convert EEG signals into graph signals having appropriate graph structures and signal features, which can overcome the low dimensionality of EEG, and use the graph convolutional neural network (GCNN) to learn the graph signals. We deal with an EEG classification problem where the visual stimulus watched by a human subject is identified through EEG.

%The rest of the paper is organized as follows. Section \ref{sec:relatedworks} reviews applications of graph signal processing to diverse brain signals. Section \ref{sec:methodology} explains the proposed approach, including graph convolutional neural network, graph construction, and feature extraction methods. Section \ref{sec:experiment} presents our experiment and detailed analysis of the results. Finally, Section \ref{sec:conclusion} concludes the paper.

% From the experiment, it was shown that the most contributing frequency signatures depend on the task familiarity.
% According to work in \cite{Rui2016dimensionality}, used brain connectivity graph constructed from correlation, coherence, phase locking value\cite{Lachaux1999measuring}, and Granger causality\cite{Brovelli2004beta} and defined signal based on MEG data. They applied graph filtering to reduce dimensionality to extract features to classify MEG signals for watching face images and non-face images.  
%comparable   proposed neural networks based on graph convolutional layer and autoencoder-like network structure. The authors conducted experiments to classify MEG signals for watching face images and non-face images.
%constructed graph from fMRI signals based on distance between cortical and subcortical region of interests (ROI). The authors designed a deep neural network consists of graph convolutional layers and fully connected layers to estimate similarity between two fMRI signals, and they applied the neural network to classify whether the signal is from a subject suffering from autism spectrum disorder or not.
\section{Related Work}
\label{sec:relatedworks}
The graph signal processing consists of merging graph theoretic concepts and signal processing concepts to extend the domains of signal processing and graph analysis \cite{Shuman2013emerging}\cite{sandryhaila2013discrete}. The brain signal has been analyzed using graph signal processing techniques in recent years. In \cite{Huang2016graph}, the authors analyzed fMRI signals of subjects during simple motor learning tasks, and extracted graph frequency signatures for different task familiarities. According to the work in \cite{Rui2016dimensionality}, applying graph filtering to reduce dimensionality of MEG signals yields better performance in classification tasks than traditional feature extraction techniques such as principal component analysis (PCA) and linear discriminant analysis (LDA).

With the recent explosion of deep learning techniques, there are studies to apply deep learning to brain signal analysis. It was shown that several deep neural networks such as deep belief networks, stacked denoising autoencoders, and convolutional neural networks can extract effective application-driven features for EEG signals in spatial and spectral domains \cite{Xu2016affective}\cite{moon2018convolutional}\cite{tabar2016novel}\cite{schirrmeister2017deep}. Deep learning models on graph signals, especially graph convolutional neural networks (GCNN) \cite{Defferrard2016convolutional}\cite{Kipf2016semi} have been considered as competitive approaches for analyzing MEG \cite{Guo2017deep} and fMRI \cite{Ktena2017distance} signals. However, graph signal-based deep learning for EEG has been rarely found in literature. A challenge arises from the fact that EEG signals usually have smaller numbers of brain region representatives, i.e., electrodes (e.g., 32 in \cite{Koelstra2012deap}), than MEG (e.g., 306 in \cite{Guo2017deep}) or fMRI (e.g., 110 in \cite{Ktena2017distance}), so it is difficult to construct rich graph structures for EEG signals. 

% Outline figure.
\setlength{\intextsep}{6pt}
\begin{figure*}[htb]
	\centering
	\begin{minipage}[b]{1.0\linewidth}
		\centering
		\centerline{\includegraphics[width=\textwidth]{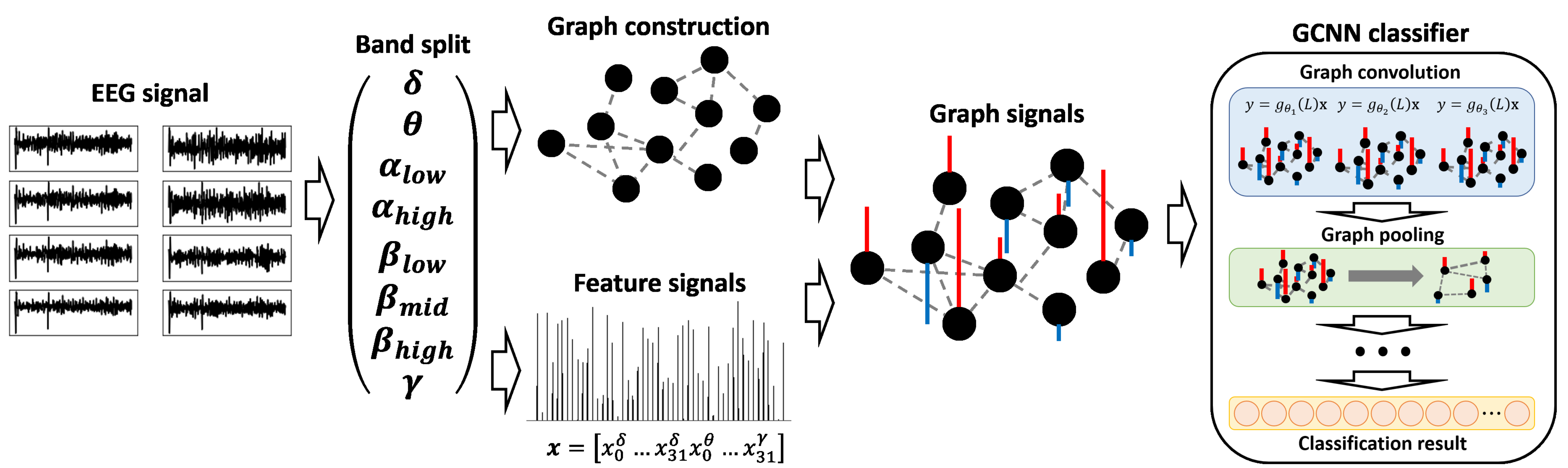}}
		%  \vspace{2.0cm}
		\medskip
	\end{minipage}
	\caption{Illustration of the proposed approach.}
	\label{fig:outline}
\end{figure*}

\section{Methodology}
\label{sec:methodology}

This section explains our approach for EEG signal classification, including extracting signal features, constructing graphs, and learning with GCNN. Figure \ref{fig:outline} illustrates our approach.

\subsection{Signal Features}
\label{sec:methodology_signal}
For analysis of EEG signals, it is common to decompose signals into multiple bands, for instance, delta (0-3Hz), theta (4-7Hz), alpha (8-12Hz), beta (13-30Hz), and gamma (31-50Hz). For more detailed analysis, the alpha and beta bands can be divided further into sub-bands to consider eight bands in total: low-alpha (8-10Hz), high-alpha (10-12Hz), low-beta (13-16Hz), mid-beta (17-20Hz), and high-beta (21-29Hz), which is employed in our work. We extract the signals of different bands using the Parks-McClellan FIR band-pass filters having an order of 47 \cite{Mcclellan2005personal}. The power and entropy of filtered signals are used as the signals on graphs \cite{Koelstra2012deap}\cite{sabeti2009entropy}.

\subsection{Graph Construction and Merging}
\label{sec:methodology_graph}
The brain consists of multiple regions separated anatomically or functionally, and the brain signal can be modeled as a graph \cite{Bullmore2009complex}. In the case of EEG, each electrode corresponds to each brain region, and the connectivity is calculated between each pair of electrodes. 

We build a graph from the given set of EEG signals from all electrodes in two steps. First, an intra-band graph is created using the signals of each band. Then, the intra-band graphs for all bands are merged to form the richer graph model.

We adopt three methods to define intra-band graphs, namely, correlation-based (corr), distance-based (dist), and random (rand) methods. First, the correlation coefficient is computed to measure the level of functional connectivity between two electrodes \cite{hassan2014eeg}, and its absolute value is used as the weight of an edge between the vertices corresponding to the electrodes. Second, the distance-based method uses the physical distances between the electrodes, which emphasizes spatial relationship between brain regions \cite{Rui2016dimensionality}. In this case, the edge connecting electrodes close to each other is assigned with a large edge weight, and vice versa. In other words, for a pair of electrodes with distance $d$, the distance connectivity is calculated by $e^{-d^2/\sigma^2}$, where $\sigma$ is a scaling constant. Third, the random method builds an Erd{\"{o}}s-R{\'{e}}nyi random graph with an edge probability of $p$. We examine this method to evaluate the importance of considering relationship between electrodes in building the graphs. In our experiment, we set $p\in\{0.3, 0.5, 0.7\}$, and all the edge weights are set to 1.

The graph obtained by correlation or distance is too densely connected to perform efficient analysis. Thus, we sparsify it to emphasize strongly connected vertices. The edge weight values for a certain vertex are sorted in a descending order, and only the top-$k$ weights are kept, while the remaining edges are removed. The value of $k$ defines the level of sparsity of the graph. We use $k\in\{4, 8, 12\}$.

Once each intra-band graph corresponding to each frequency band is obtained, we have eight sets of signal features and graphs for one signal segment. We consider two methods to merge them and obtain a larger graph. In the first method, we simply collect the intra-band graphs to compose the final graph having eight components. Thus, there is no edge between intra-band graphs. In the second method, we impose inter-band connectivity between all pairs of the vertices corresponding to the same electrodes but different bands, and the weights of all these edges are set to 1. Hence, we obtain the final graph containing both intra-band and inter-band connectivity.

\subsection{Graph Convolutional Neural Network (GCNN)}
\label{sec:methodology_gcnn}
The graph convolutional neural network (GCNN) \cite{Defferrard2016convolutional}, also known as ChebNet, performs convolution and pooling for signals defined on graphs based on graph signal processing and graph coarsening.

For an undirected and weighted graph $G=\{V,E,W\}$ with $N$ vertices, where $V$ is a set of vertices, $E$ is a set of edges, and $W$ is an $N{\times}N$ real matrix containing edge weights. In graph signal processing, a signal $x:V\to\mathbb{R}$ is defined on graph vertices, and can be represented by a vector $\mathbf{x}\in\mathbb{R}^N$ when all the $N$ vertices are considered. The normalized graph Laplacian is defined as $\tilde{L}=I_{N}-D^{-1/2}W D^{-1/2}$, where $I_N$ is an identity matrix with size $N$ and $D$ is the diagonal matrix with degree of each vertex. Note that the eigenvalues of $\tilde{L}$ are within $[0,2]$. The spectrum of the normalized graph Laplacian is defined from its eigendecomposition, $\tilde{L}=U\Lambda U^T$, where the columns of $U$ consist of eigenvectors $\{u_l\}, l=1...N$, and $\Lambda$ is a diagonal matrix that has graph frequencies $\{\lambda_l\}, l=1...N$, as the diagonal elements. Then, the graph Fourier transform of graph signal $\mathbf{x}$ on graph $G$ is defined as $\tilde{\mathbf{x}}=U^{T}\mathbf{x}$, and the inverse graph Fourier transform is defined as $\mathbf{x}=U\tilde{\boldmath{x}}$.

Convolution of signal $\mathbf{x}$ and filter $\theta$ on the graph in GCNN is executed on the spectral domain:
\begin{equation}
\label{eq:convolution}
y=g_\theta(L)\mathbf{x}=U g_\theta(\Lambda)U^{T}\mathbf{x}
\end{equation}
where $g_\theta$ is a graph signal filtering function. To achieve localized filtering, the filter must be a polynomial of the graph Laplacian, i.e., $g_\theta(\Lambda)=\sum_{m=0}^{M-1}\theta_m\Lambda^m$, where $M$ is the order of a filtering polynomial and $\theta_{m}$ is the coefficient for polynomial order $m$. In GCNN, the Chebyshev polynomial is used to reduce calculation of spectral filtering. Therefore, the filter is defined as
\begin{equation}
\label{eq:chebyshevfilter}
g_\theta(\Lambda)=\sum_{m=0}^{M-1}\theta_{m}T_{m}(\tilde{\Lambda}) 
\end{equation}
where $T_{m}(\tilde{\Lambda})$ is the Chebyshev polynomial of order $m$, evaluated at $\tilde{\Lambda}$. Here, $\tilde{\Lambda}$ is a diagonal matrix of scaled eigenvalues defined as $\tilde{\Lambda}=2\Lambda /\lambda_{max} - I_N$.

A graph pooling layer in GCNN pools information from multiple vertices to one vertex, to reduce the graph and to expand receptive fields of graph signal filters. The graph pooling is implemented based on the graph coarsening algorithm \cite{dhillon2007weighted}, with a fake vertex trick to construct binary trees for fast pooling. The feature vectors from the last graph convolutional layer are concatenated into a single feature vector, which is fed into the fully connected layer to obtain classification results.

\section{Experiment}
\label{sec:experiment}

We conduct an EEG-based video identification experiment to evaluate the proposed approach.

\subsection{Dataset}
\label{sec:experiment_dataset}
The DEAP \cite{Koelstra2012deap} is one of the largest datasets containing human affective states. It contains 32-channel EEG signals of 32 participants, recorded during watching 40 music video stimuli. We use the preprocessed version of the dataset, which has a sampling rate of 128Hz.

The EEG signals of a subject for a video consists of three-second pre-trial baseline signals and 60-second trial signals. The trial signals are divided into three-second-long segments with a stride of one second. Then, the power and entropy values are calculated for the eight frequency bands. As a result, we obtain 58 samples of 32$\times$8-dimensional feature vectors for each recording.

\subsection{Experiment Design}
\label{sec:experiment_design}

We implement the experiment using Tensorflow \cite{Abadi2016tensorflow}. The GCNN network has four graph convolutional layers, two graph pooling layers, and a fully connected layer, which is denoted as:
\begin{itemize}
	\item GC64M16 - GC64M16 - P2 - GC128M9 - GC128M9 - P2 - FC40
\end{itemize}
where GC$(f)$M$(m)$ means a graph convolutional layer with $f$ filters and polynomial order of $m$, P$(c)$ means a pooling layer that reduces the number of vertices by a factor of $c$, and FC$(h)$ corresponds to a fully connected layer with $h$ output neurons. The network is trained using the Adam \cite{Kingma2014adam} optimizer to minimize the cross-entropy loss with L2-regularization for 30 epochs. We set the initial learning rate to 0.001 and it decreases with a decay rate of 0.95 at every epoch. We use 80\% of signal segments for training, and the others for test.

\subsection{Results}
\label{sec:experiment_result}

\setlength{\textfloatsep}{10pt plus 2pt minus 4pt}
\setlength{\intextsep}{10pt plus 4pt minus 4pt}
\begin{table}[!b]
	\centering
	\caption{Accuracy (\%) of video identification with GCNNs and baseline classifiers.}
	\label{table:net2_result}
	\begin{tabular}{|c|c|l|r|r|}
		\hline
		\multirow{2}{*}{Graph}  & \multirow{2}{*}{Inter-band} & \multicolumn{1}{c|}{\multirow{2}{*}{Density}} & \multicolumn{2}{c|}{Signal}                               \\ \cline{4-5} 
		&                          & \multicolumn{1}{c|}{}                         & \multicolumn{1}{c|}{Power} & \multicolumn{1}{c|}{Entropy} \\ \hline \hline
		\multirow{6}{*}{corr}    & \multirow{3}{*}{$\times$} & $k=$4    & 56.56   & 57.30                        \\ \cline{3-5} 
		&                                                    & $k=$8    & 56.40   & 57.09                        \\ \cline{3-5} 
		&                                                    & $k=$12   & 57.80   & 56.72                        \\ \cline{2-5}
		& \multirow{3}{*}{$\bigcirc$}                        & $k=$4    & 61.94   & 65.25                        \\ \cline{3-5} 
		&                                                    & $k=$8    & 61.74   & 62.53                        \\ \cline{3-5} 
		&                                                    & $k=$12   & 61.20   & 61.57                        \\ \hline
		\multirow{6}{*}{dist}   & \multirow{3}{*}{$\times$}  & $k=$4    & 58.49   & 56.67                        \\ \cline{3-5} 
		&                                                    & $k=$8    & 57.00   & 54.76                        \\ \cline{3-5} 
		&                                                    & $k=$12   & 58.57   & 58.40                        \\ \cline{2-5} 
		& \multirow{3}{*}{$\bigcirc$}                        & $k=$4    & 64.15   & 65.27                        \\ \cline{3-5} 
		&                                                    & $k=$8    & 62.82   & 62.65                        \\ \cline{3-5} 
		&                                                    & $k=$12   & 61.92   & 62.13                        \\ \hline
		\multirow{6}{*}{rand} & \multirow{3}{*}{$\times$}    & $p=$0.3  & 57.12   & 56.81                        \\ \cline{3-5} 
		&                                                    & $p=$0.5  & 58.57   & 57.35                        \\ \cline{3-5} 
		&                                                    & $p=$0.7  & 58.69   & 55.88                        \\ \cline{2-5} 
		& \multirow{3}{*}{$\bigcirc$}                        & $p=$0.3  & 62.90   & 63.19                        \\ \cline{3-5} 
		&                                                    & $p=$0.5  & 61.70   & 62.03                        \\ \cline{3-5} 
		&                                                    & $p=$0.7  & 59.61   & 57.67                        \\ \hline
		\multicolumn{3}{|c|}{k-nearest neighbors}                       & 40.46   & 48.50                        \\ \hline
		\multicolumn{3}{|c|}{random forest}                             & 51.34   & 42.60                        \\ \hline
	\end{tabular}
\end{table}

Table \ref{table:net2_result} presents the results of the proposed method. We compare our results with a k-nearest neighbor classifier and a random forest classifier as baselines. It is observed that the proposed method with GCNN shows significantly better performance than the baseline methods overall. The highest accuracy by GCNN is 65.27\%, whereas the two baseline methods produce accuracies of 48.50\% and 51.34\%. Thus, involving graph structures for classification of EEG helps extracting meaningful patterns.

The performance is mainly affected by the graph construction method. Especially, the graphs with inter-band connections yield better performance than those without inter-band connections. The inter-band connections assign global connectivity to the overall graph, which facilitates extracting useful representations between multiple frequency bands. 

When the three types of intra-band graphs are compared, the graphs constructed using correlation and distance are similarly better than the random graphs, showing that the elaborate graph construction is advantageous. Regarding the density of the intra-band graphs (controlled by $k$ for dist and corr, or $p$ for rand), higher accuracies are obtained by lower density values. This result indicates that the excessive complexity of a graph due to too many edges is not beneficial. The entropy as the signal feature shows slightly better performance than the power for low-density graphs.

We evaluate various structures of GCNN to test relationship between model complexity and classification performance. We test five network structures as follows:

\noindent
\setlist{nolistsep}
\setlength{\textfloatsep}{20.4pt plus 2.4pt minus 4.8pt}
\setlength{\intextsep}{12.0pt plus 2.4pt minus 2.4pt}
\begin{itemize}
	\itemsep0em 
	\item \textit{Network} 1: GC64M16 - GC64M16 - P2 - FC40
	\item \textit{Network} 2: GC64M16 - GC64M16 - P2 - GC128M9 - GC128M9 - P2 - FC40
	\item \textit{Network} 3: GC64M9 - GC64M9 - P2 - GC128M4 - GC128M4 - P2 - FC40
	\item \textit{Network} 4: GC64M4 - GC64M4 - P2 - GC128M3 - GC128M3 - P2 - FC40
	\item \textit{Network} 5: GC64M16 - GC64M16 - P2 - GC128M9 - GC128M9 - P2 - GC256M4 - GC256M4 - P2 - FC40
	\\
\end{itemize}

%
%\begin{minipage}{\linewidth}
%
%\end{minipage}

%\begin{itemize}
%	\item \textit{Network} 1: GC64M16 - GC64M16 - P2 - FC40
%	\item \textit{Network} 2: GC64M16 - GC64M16 - P2 - GC128M9 - GC128M9 - P2 - FC40
%	\item \textit{Network} 3: GC64M16 - GC64M16 - P2 - GC128M9 - GC128M9 - P2 - GC256M4 - GC256M4 - P2 - FC40
%\end{itemize}
Note that \textit{Network} 2 is the one used in Table \ref{table:net2_result}, and \textit{Network 3} and \textit{Network} 4 are the same to \textit{Network} 2 except for reduced polynomial orders. In this experiment, the best combination of graph and signal construction methods in the previous evaluation are selected, i.e., the graph is created using the distance-based intra-band graph connectivity for $k=4$, with enabling inter-band connections, and the entropy is used for the signal feature. The results of evaluation are presented in Table \ref{table:networkperformance}. \textit{Network} 2 with 4 graph convolutional layers performs better than \textit{Network} 1 or \textit{Network} 5, showing that too simple or too complex a network produces performance degradation. \textit{Network} 3 and \textit{Network} 4 are out performed by \textit{Network} 2, indicating that a sufficiently large polynomial order is also important for classification performance, which is related to the receptive field size of a graph convolutional filter. Interestingly, when \textit{Network} 1 and \textit{Network} 4, which have similar numbers of parameters, are compared, the depth of GCNN seems more important than the polynomial order. This may be analogous to the trend of "going deep" using small filters in CNNs \cite{simonyan2014very}. 

\setlength{\textfloatsep}{10pt plus 2pt minus 4pt}
\setlength{\intextsep}{10pt plus 4pt minus 4pt}
\begin{table}[!h]
	\small%
	\centering
	\caption{\label{table:networkperformance} Accuracy of video identification for various GCNN structures.}
	\centering
	\begin{tabular}[pos=b]{|c|c|c|} \hline
		Network type & \#parameter & Accuracy \\\hline \hline
		\textit{Network} 1& 722k & 48.25\% \\\hline
		\textit{Network} 2& 944k & 65.27\% \\\hline
		\textit{Network} 3& 792k & 57.41\% \\\hline
		\textit{Network} 4& 746k & 51.83\% \\\hline
		\textit{Network} 5& 1337k & 62.94\% \\\hline
	\end{tabular}
\end{table}

%\begin{figure}
%	\centering
%	\begin{subfigure}{.25\textwidth}
%		\centering
%		\includegraphics[width=\linewidth]{confusion_matrix}
%		\caption{ }
%		\label{fig:sub_confusion}
%	\end{subfigure}%
%	\begin{subfigure}{.25\textwidth}
%		\centering
%		\includegraphics[width=\linewidth]{video_accuracy}
%		\caption{ }
%		\label{fig:sub_accuracy}
%	\end{subfigure}
%	\caption{Confusion matrix (a) and accuracies (b) of the experiment using \textit{Network} 2 classifier and distance-based intra-band graph connectivity for $k$=4 with enabling inter-band connections, and entropy signals.}
%	\label{fig:result_analysis}
%\end{figure}
%
%Figure \ref{fig:result_analysis} shows confusion matrix and accuracies from the experiment. From two-tailed z-tests, there are 11 videos which have significantly different accuracy from the average (65.27\%) at a significance level of 0.01, but these videos do not have common features in popularity or emotion responses by subjects, including valence, arousal, and dominance.

\section{Conclusion}
\label{sec:conclusion}

We have proposed a EEG classification approach using GCNN, particularly the methods for graph construction and graph signal feature extraction. It was shown that constructing graphs by extending vertices using intra-band and inter-band connectivity brought considerable performance gain in video identification. Especially, the inter-band connectivity made the graph richer to exploit the relationship between different frequency bands, which led to wider receptive fields over graphs representing in the brain. In the future, the relationship between graph construction schemes and signal features will be studied more extensively.

\bibliographystyle{IEEEbib}
\bibliography{refs}

\begin{thebibliography}{10}

\bibitem{van2010exploring}
M.~P.~Van den Heuvel and H.~E.~Hulshoff Pol,
\newblock ``Exploring the brain network: a review on resting-state {fMRI}
  functional connectivity,''
\newblock {\em European {N}europsychopharmacology}, vol. 20, no. 8, pp.
  519--534, 2010.

\bibitem{hassan2014eeg}
M.~Hassan, O.~Dufor, I.~Merlet, C.~Berrou, and F.~Wendling,
\newblock ``{EEG} source connectivity analysis: from dense array recordings to
  brain networks,''
\newblock {\em PLoS ONE}, vol. 9, no. 8, pp. e105041, 2014.

\bibitem{Shuman2013emerging}
D.~I. Shuman, S.~K. Narang, P.~Frossard, A.~Ortega, and P.~Vandergheynst,
\newblock ``The emerging field of signal processing on graphs: Extending
  high-dimensional data analysis to networks and other irregular domains,''
\newblock {\em IEEE Signal Processing Magazine}, vol. 30, no. 3, pp. 83--98,
  2013.

\bibitem{sandryhaila2013discrete}
A.~Sandryhaila and J.~M.~F. Moura,
\newblock ``Discrete signal processing on graphs,''
\newblock {\em IEEE Transactions on Signal Processing}, vol. 61, no. 7, pp.
  1644--1656, 2013.

\bibitem{Defferrard2016convolutional}
M.~Defferrard, X.~Bresson, and P.~Vandergheynst,
\newblock ``Convolutional neural networks on graphs with fast localized
  spectral filtering,''
\newblock in {\em Advances in Neural Information Processing Systems}, 2016, pp.
  3844--3852.

\bibitem{Kipf2016semi}
T.~N. Kipf and M.~Welling,
\newblock ``Semi-supervised classification with graph convolutional networks,''
\newblock {\em arXiv preprint arXiv:1609.02907}, 2016.

\bibitem{Pham2017column}
T.~Pham, T.~Tran, D.~Q. Phung, and S.~Venkatesh,
\newblock ``Column networks for collective classification,''
\newblock in {\em Proceedings of AAAI Conference on Artificial Inteligence},
  2017, pp. 2485--2491.

\bibitem{Huang2016graph}
W.~Huang, L.~Goldsberry, N.~F. Wymbs, S.~T. Grafton, D.~S. Bassett, and
  A.~Ribeiro,
\newblock ``Graph frequency analysis of brain signals,''
\newblock {\em IEEE Journal of Selected Topics in Signal Processing}, vol. 10,
  no. 7, pp. 1189--1203, 2016.

\bibitem{Rui2016dimensionality}
L.~Rui, H.~Nejati, and N.-M. Cheung,
\newblock ``Dimensionality reduction of brain imaging data using graph signal
  processing,''
\newblock in {\em Proceedings of International Conference on Image Processing},
  2016, pp. 1329--1333.

\bibitem{Xu2016affective}
H.~Xu and K.~N. Plataniotis,
\newblock ``Affective states classification using {EEG} and semi-supervised
  deep learning approaches,''
\newblock in {\em Proceedings of IEEE International Workshop on Multimedia
  Signal Processing}, 2016, pp. 1--6.

\bibitem{moon2018convolutional}
S.-E. Moon, S.~Jang, and J.-S. Lee,
\newblock ``Convolutional neural network approach for {EEG}-based emotion
  recognition using brain connectivity and its spatial information,''
\newblock in {\em Proceedings of IEEE International Conference on Acoustics,
  Speech and Signal Processing}, 2018.

\bibitem{tabar2016novel}
Y.~R. Tabar and U.~Halici,
\newblock ``A novel deep learning approach for classification of {EEG} motor
  imagery signals,''
\newblock {\em Journal of Neural Engineering}, vol. 14, no. 1, pp. 016003,
  2016.

\bibitem{schirrmeister2017deep}
R.~T. Schirrmeister, J.~T. Springenberg, L.~D.~J. Fiederer, M.~Glasstetter,
  K.~Eggensperger, M.~Tangermann, F.~Hutter, W.~Burgard, and T.~Ball,
\newblock ``Deep learning with convolutional neural networks for {EEG} decoding
  and visualization,''
\newblock {\em Human Brain Mapping}, vol. 38, no. 11, pp. 5391--5420, 2017.

\bibitem{Guo2017deep}
Y.~Guo, H.~Nejati, and N.-M. Cheung,
\newblock ``Deep neural networks on graph signals for brain imaging analysis,''
\newblock {\em arXiv preprint arXiv:1705.04828}, 2017.

\bibitem{Ktena2017distance}
S.~I. Ktena, S.~Parisot, E.~Ferrante, M.~Rajchl, M.~Lee, B.~Glocker, and
  D.~Rueckert,
\newblock ``Distance metric learning using graph convolutional networks:
  Application to functional brain networks,''
\newblock {\em arXiv preprint arXiv:1703.02161}, 2017.

\bibitem{Koelstra2012deap}
S.~Koelstra, C.~M{\"{u}}hl, M.~Soleymani, J.-S. Lee, A.~Yazdani, T.~Ebrahimi,
  T.~Pun, A.~Nijholt, and I.~Patras,
\newblock ``{DEAP}: A database for emotion analysis; using physiological
  signals,''
\newblock {\em IEEE Transactions on Affective Computing}, vol. 3, no. 1, pp.
  18--31, 2012.

\bibitem{Mcclellan2005personal}
J.~H. McClellan and T.~W. Parks,
\newblock ``A personal history of the {P}arks-{M}c{C}lellan algorithm,''
\newblock {\em IEEE Signal Processing Magazine}, vol. 22, no. 2, pp. 82--86,
  2005.

\bibitem{sabeti2009entropy}
M.~Sabeti, S.~Katebi, and R.~Boostani,
\newblock ``Entropy and complexity measures for {EEG} signal classification of
  schizophrenic and control participants,''
\newblock {\em Artificial Intelligence in Medicine}, vol. 47, no. 3, pp.
  263--274, 2009.

\bibitem{Bullmore2009complex}
E.~Bullmore and O.~Sporns,
\newblock ``Complex brain networks: graph theoretical analysis of structural
  and functional systems,''
\newblock {\em Nature {R}eviews {N}euroscience}, vol. 10, no. 3, pp. 186--198,
  2009.

\bibitem{dhillon2007weighted}
I.~S. Dhillon, Y.~Guan, and B.~Kulis,
\newblock ``Weighted graph cuts without eigenvectors a multilevel approach,''
\newblock {\em IEEE Transactions on Pattern Analysis and Machine Intelligence},
  vol. 29, no. 11, pp. 1944--1957, 2007.

\bibitem{Abadi2016tensorflow}
M.~Abadi et~al.,
\newblock ``{T}ensorflow: A system for large-scale machine learning,''
\newblock in {\em Proceedings of the USENIX Symposium on Operating Systems
  Design and Implementation}, 2016, vol.~16, pp. 265--283.

\bibitem{Kingma2014adam}
D.~Kingma and J.~Ba,
\newblock ``Adam: A method for stochastic optimization,''
\newblock {\em arXiv preprint arXiv:1412.6980}, 2014.

\bibitem{simonyan2014very}
K.~Simonyan and A.~Zisserman,
\newblock ``Very deep convolutional networks for large-scale image
  recognition,''
\newblock {\em arXiv preprint arXiv:1409.1556}, 2014.

\end{thebibliography}

\end{document}